\documentclass{aastex}
\usepackage{emulateapj5,apjfonts,epsfig}

\slugcomment{To appear in The Astrophysical Journal, Vol. 573 (2002 July 10)}

\shorttitle{OAO 1657$-$415}
\shortauthors{CHAKRABARTY ET AL.}

\begin{document}

\title{The X-Ray Position and Infrared Counterpart of the Eclipsing
X-Ray Pulsar OAO~1657$-$415}

\author{Deepto~Chakrabarty,\altaffilmark{1,2}
   Zhongxiang Wang,\altaffilmark{1} Adrienne~M.~Juett, and Julia~C.~Lee}
\affil{Department of Physics and Center for Space Research, Massachusetts
   Institute of Technology, Cambridge, MA 02139; 
   deepto@space.mit.edu, wangzx@space.mit.edu, ajuett@space.mit.edu, 
   jlee@space.mit.edu}

\medskip
\and

\author{Paul~Roche\altaffilmark{3}}
\affil{Department of Physics and Astronomy, University of Leicester,
   Leicester LE1 7RH, UK; pdr@star.le.ac.uk} 

\altaffiltext{1}{Visiting astronomer, Cerro Tololo Inter-American
   Observatory, National Optical Astronomy Observatories, operated by
   the Association of Universities for Research in Astronomy under
   contract to the National Science Foundation.} 

\altaffiltext{2}{Alfred P. Sloan Research Fellow}

\altaffiltext{3}{Also Earth and Space Sciences, School of Applied
   Sciences, University of Glamorgan, Pontypridd CF37 1DL, Wales, UK}

\begin{abstract}
We have measured the precise position of the 38 s eclipsing X-ray
pulsar OAO~1657$-$415 with the {\em Chandra X-Ray Observatory}:
$\alpha$(J2000) = 17$^{\rm h}$00$^{\rm m}$48\fs90, $\delta$(J2000)
= $-41^\circ$39\arcmin21\farcs6, error radius = 0\farcs5.  Based on
the previously measured pulsar mass function and X-ray eclipse
duration, this 10.4-d high-mass X-ray binary is believed to contain a
B~supergiant companion.  Deep optical imaging of the field did not
detect any stars at the {\em Chandra} source position, setting a limit
of $V>23$.  However, near-infrared imaging revealed a relatively
bright star ($J=14.1$, $H=11.9$, $K_s=10.7$) coincident with the {\em
Chandra} position, and we identify this star as the infrared 
counterpart of OAO~1657$-$415.  The infrared colors and magnitudes and the
optical non-detections for this star are all consistent with a highly
reddened B supergiant ($A_V=20.4\pm 1.3$) at a distance of $6.4\pm
1.5$~kpc.  This implies an X-ray luminosity of $3\times 10^{36}$
erg~s$^{-1}$ (2--10 keV).  Infrared spectroscopy can verify the
spectral type of the companion and measure its radial velocity curve,
yielding a neutron star mass measurement.
\end{abstract}


\section{INTRODUCTION}

Nearly a decade ago, observations with the {\em Compton}/BATSE all-sky
monitor revealed that the 38~s accretion-powered X-ray pulsar
OAO~1657$-$415 ($l=344^\circ$, $b=$0\fdg3) is in a 10.4~d eclipsing
binary with an unidentified B~supergiant companion (Chakrabarty et
al. 1993).  The nature of the mass donor was inferred from X-ray
timing measurements and the X-ray eclipse duration.  Only six other
eclipsing X-ray pulsars are known, and all of them yield important
constraints on the neutron star mass range (van Kerkwijk, van
Paradijs, \& Zuiderwijk 1995; Chakrabarty, Psaltis, \& Thorsett 2002,
in preparation).  Although binary radio pulsar data yield more precise
neutron star mass measurements (Thorsett \& Chakrabarty 1997), the
X-ray binaries generally trace a different evolutionary path and
therefore may have a systematically different mass range.

OAO~1657$-$415 is also unique among the known high-mass X-ray binaries
in that it appears to occupy a transition region between mass transfer
via a stellar wind and Roche-lobe overflow, with possible episodic
formation of an accretion disk (Chakrabarty et al. 1993; Bildsten et
al. 1997; Baykal 1997, 2000).  Since the binary is too wide for Roche
lobe overflow to occur, this may provide the first clear evidence that
the winds in high-mass X-ray binaries possess sufficient angular
momentum to form accretion disks.  Identification of the supergiant
companion and follow-up spectroscopy would permit both a neutron star
mass measurement as well as a search for accretion disk signatures.

However, optical identification has been hampered by the source's
poorly known X-ray position.  The most precise previous measurement
was derived from a 0.5--4.0~keV {\em Einstein}/IPC image (Parmar et
al. 1980).  A reanalysis of these archival data yields R.A. = 17$^{\rm
h}$00$^{\rm m}$47\fs70 and decl. = $-41^\circ$39\arcmin15\farcs5
(equinox J2000.0), with a 90\%-confidence error radius of 32\arcsec
(Harris et al. 1994).  The source is very heavily absorbed ($N_{\rm
H}\sim 10^{23}$ cm$^{-2}$; Polidan et al. 1978; Parmar et al. 1980;
Kamata et al. 1990), preventing refined localization with {\em ROSAT}.
Indeed, an 11~ks {\em ROSAT}/HRI (0.2--2.4~keV) observation failed to
detect the source at all (Chakrabarty 1995, unpublished).  Previous
optical work has shown that there are no OB supergiants in the {\em
Einstein} error circle with magnitude $V<19$ (Roche 1993; Maxwell,
Norton, \& Roche 2001), indicating that the luminous optical
counterpart is also subject to heavy extinction.

In this Letter, we report a precise position measurement of OAO~1657$-$415
using the {\em Chandra X-Ray Observatory} and the identification of an
infrared counterpart, presumably an OB supergiant.  We discuss our
imaging observations (X-ray, optical, and infrared) in \S2, our
X-ray spectroscopy in \S3, and the implications of our observations in
\S4. 

\section{X-RAY AND OPTICAL/IR IMAGING}

We observed OAO 1657$-$415 with {\em Chandra} on 2001 February 10 for
5.1~ks using the High Energy Transmission Grating Spectrometer (HETGS)
and the spectroscopy array of the Advanced CCD Imaging Spectrometer
(ACIS-S).  The HETGS employs two sets of transmission gratings: the
Medium Energy Gratings (MEGs; 0.4--5.0~keV) and the High Energy
Gratings (HEGs; 0.8--10.0~keV).  The HETGS spectra were imaged by
ACIS-S, an array of six CCD detectors.  The HETGS/ACIS-S combination
provides an undispersed (zeroth-order) image 
\centerline{\epsfxsize=8.5cm\epsfbox{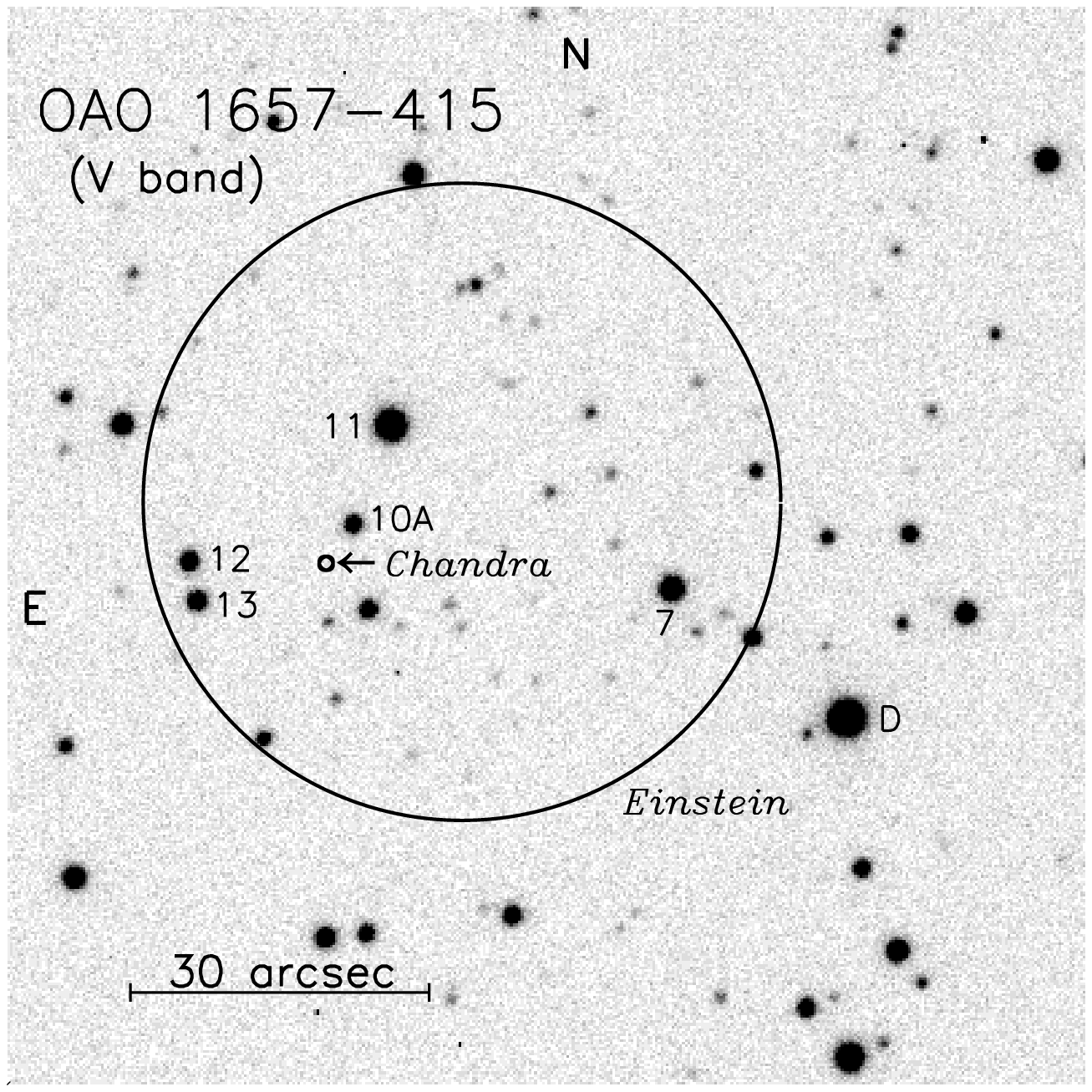}}
\figcaption{$V$-band finder for the OAO 1657$-$415 field.  The
32\arcsec\ {\em Einstein} error circle and the 0\farcs5 {\em Chandra}
error circle are indicated.  No star is detected at the {\em Chandra}
position down to a limiting magnitude of 23.1.  For reference, several
field stars are labeled using the numbering scheme of Roche (1993).
Star~D of Maxwell et al. (2001) is also noted.}

\bigskip
\noindent
and dispersed spectra from the gratings.  The various orders overlap
and are sorted using the intrinsic energy resolution of the ACIS CCDs,
which are read out every 3.2~s.  Pulsations were detected from
OAO~1657$-$415 at a barycentered period of 37.329$\pm$0.020~s (epoch
MJD 51950.84), verifying that the detected source is the pulsar.
Besides the pulsar, no other X-ray sources are detected in the {\em
Chandra} image.

The zeroth order HETGS image of OAO 1657$-$415 was affected by photon
pileup (see, e.g., Davis 2001), but not so severely as to distort the
image centroid.  We measured the source position with the CIAO tool
{\em celldetect}.\footnote{See 
http://cxc.harvard.edu/ciao.}  Our best-fit position was R.A. =
17$^{\rm h}$00$^{\rm m}$48\fs90 and decl. =
$-41^\circ$39\arcmin21\farcs6 (equinox J2000.0), with an approximate
error radius of 0\farcs5 (Aldcroft et al. 2000\footnote{ See also 
http://cxc.harvard.edu/cal/ASPECT/celmon/.}).  This position lies
within the 32\arcsec\ {\em Einstein} error circle but is 15\arcsec\
from its center.  The {\em Chandra} position excludes the candidate
optical counterpart suggested by Maxwell, Norton, \& Roche (2001;
star~D in Figure~1), which lies 55\arcsec\ away.

We obtained deep {\em UBV}-band optical images of the OAO 1657$-$415
field on 1999 August~8 using the EMMI camera at the f/11 Nasmyth~B
focus of the 3.5-m New Technology Telescope (NTT) at the European
Southern Observatory (ESO) at Cerro La Silla, Chile.  These
observations were made through the NTT service observing program.  We
derived an astrometric solution for these images by matching 21 field
stars to the USNO-A2.0 catalog of astrometric standards (Monet et
al. 1998).  The rms error in the positions was 0\farcs25.  Applying
this solution, we found no stars at the {\em Chandra} position for
OAO~1657$-$415.  Limits on the brightness of the optical counterpart
are given in Table~1.  The flux calibration of these images was done
by comparison with observations of photometric standards (Landolt
1992).  We show a $V$ image of the field in Figure~1

More recently, we obtained {\em JHK$_s$} near-infrared images of the 
\centerline{\epsfxsize=8.5cm\epsfbox{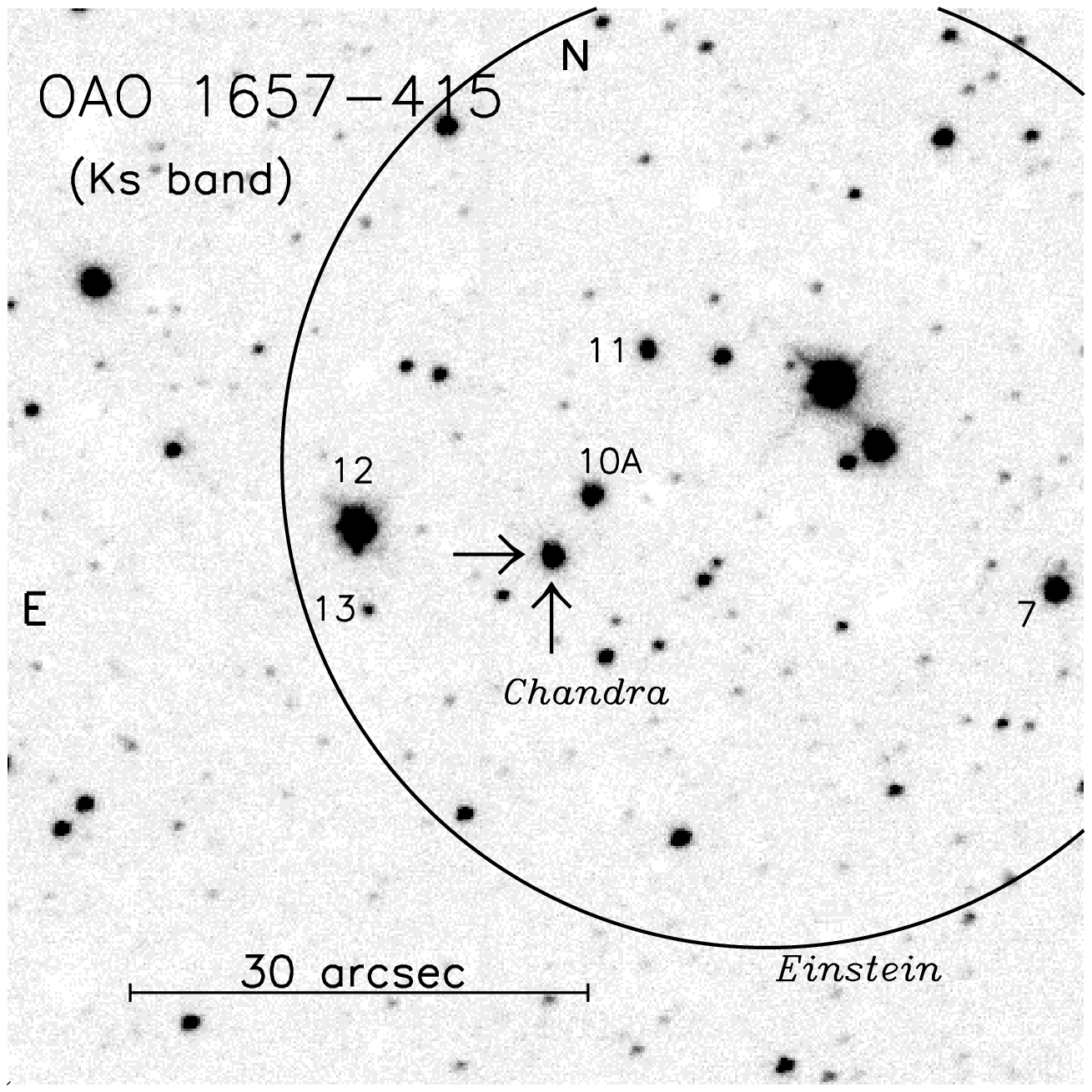}}
\figcaption{$K_s$-band finder for the OAO 1657$-$415 field.  The
infrared counterpart is indicated by the arrows.  For reference, the
32\arcsec\ {\em Einstein} error circle is shown and several of the
field stars are labeled as in Figure~2.  Some of the bright field stars
in the $K_s$ image are undetected in the $V$ image, indicating how
heavily absorbed this field is.} 

\bigskip
\noindent
field on 2002 February~25 using the Ohio State Infrared
Imager/Spectrometer (OSIRIS) at the f/14 tip-tilt focus of the 4-m 
Blanco Telescope at the Cerro Tololo Inter-American Observatory (CTIO)
in Chile.  We derived an astrometric solution for these images by
matching 12 stars in the $K_s$ image with the $V$ image described
above.  (The field of view was too small to allow a direct tie to the
USNO-A2.0 catalog.)  The rms error in the positions was dominated by
the optical fit errors.  Using this solution, we found a relatively
bright star within 0\farcs1 of the {\em Chandra} position; within the
position uncertainties, these positions are coincident, and we
identify this star as the infrared counterpart of OAO~1657$-$415.
(This object is called star~10B in the infrared images of Roche 1993.)
A $K_s$ image of the field is shown in Figure~2.  We flux-calibrated
our infrared images by comparison with observations of photometric
standards at various airmasses (Persson et al. 1998).  The measured
{\em JHK$_s$} magnitudes are given in Table~1, and are similar to
those measured by Roche (1993).

\section{X-RAY SPECTROSCOPY}

Our short {\em Chandra}/HETGS observation also yielded an X-ray
spectrum.  The combined $+/-$ first-order dispersed MEG and HEG spectra were
extracted and simultaneously fitted to an absorbed power-law $+$
gaussian model.  The observed HEG count spectrum is shown in Figure~3.
Very few counts were detected below 4 keV due to the
large hydrogen column density of $N_{\rm H} \approx 4\times 10^{23}$
cm$^{-2}$ found by fitting the absorption model to the spectra.  The
power-law index and normalization are poorly determined due to the
limited energy range available for fitting (3.5--9.0 keV).  An Fe
K$\alpha$ emission line is clearly detected at 6.4 keV, with flux
$\approx 8\times 10^{-4}$ photon cm$^{-2}$ s$^{-1}$ and equivalent
width $\approx$111~eV (we do not quote formal uncertainties, as the 
line strength is poorly constrained due to the large uncertainty 
\centerline{\epsfxsize=8.5cm\epsfbox{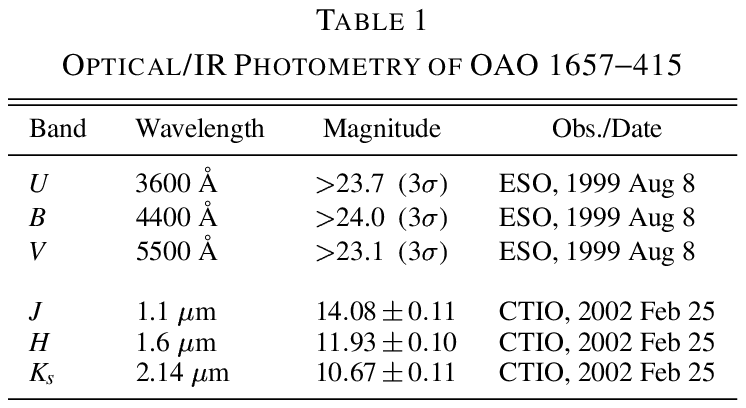}}

\bigskip
\noindent
in the continuum level).  Additionally, we detect the Fe K
photoelectric absorption edge near 7.1~keV.  Based on the strength 
of this edge, we estimate an optical depth $\tau \sim 1.2$ which
implies an even higher hydrogen column density of $N_{\rm H}\sim
10^{24}$ cm$^{-2}$ for the solar abundance values of Wilms, Allen, \&
McCray (2000).  However, given the limited statistics of our short
observation, this discrepancy is not serious.  The total observed flux
(which is not sensitive to the detailed spectral model) was $1.9\times
10^{-10}$ erg~cm$^{-2}$~s$^{-1}$ (2--10 keV).  For photon index
$\Gamma=1$, this corresponds to an unabsorbed flux of $6.4\times
10^{-10}$ erg~cm$^{-2}$~s$^{-1}$ (2--10 keV).

\section{DISCUSSION}

We have precisely measured the X-ray position of OAO~1657$-415$ and
have identified its infrared counterpart.  Based on the X-ray pulsar's
orbital parameters and the duration of the X-ray eclipse, Chakrabarty
et al. (1993) deduced that the mass donor in this binary has a mass of
14--18~$M_\odot$ and a radius of 25--32~$R_\odot$, corresponding to a
B0--6 supergiant.  They additionally estimated that the source
distance is $\gtrsim 11$~kpc using the pulsar's accretion torque behavior.   
We can compare these predictions with the observed properties of the
counterpart.  The intrinsic infrared colors of a B0--6 supergiant are
$J-H= -0.02 \pm 0.04$ and $H-K=-0.04 \pm 0.04$ (Whittet \& van Breda
1980).  Using the interstellar reddening relation of Rieke \& Lebofsky
(1985), the observed colors imply a very large reddening of
$A_V=20.4\pm 1.3$.  

We can also estimate the source distance.  For the inferred mass and
radius range, the companion should have a luminosity of $\log
(L/L_\odot)=4.8\pm 0.2$ and a temperature of $\log T =4.28\pm 0.08$
(Maeder \& Meynet 1989).  Applying the appropriate bolometric
corrections (e.g., Drilling \& Landolt 2000), the expected absolute
visual magnitude of the star is $M_V= -6.0\pm 0.2$.  From the
intrinsic B supergiant color $V-J = -0.3 \pm 0.2$ (Whittet \& van
Breda 1980) and our observed $J$ magnitude, we thus derive a source
distance of $D=6.4\pm 1.5$~kpc, somewhat closer than estimated from
the X-ray pulsar's spin-up rate.  The $H$ and $K_s$ magnitudes yield
very similar distance values.  The observed optical limits are also
consistent with our derived reddening and distance.  

Thus, we conclude that the photometric properties of the infrared
counterpart are consistent with the B supergiant companion predicted
by Chakrabarty et al. (1993).  We note, however, that the infrared
photometry alone cannot reliably provide a spectral classifiction for
the companion, since the reddening and spectral type are degenerate on
an infrared color-color diagram.  However, infrared spectroscopy can
provide an accurate spectral classification of the counterpart
(Hanson, Conti, \& Rieke 1996; Blum et al. 1997; Wallace et al. 2000)
and should eventually allow a radial velocity curve to be
measured.\ \ \ \ This

\centerline{\epsfxsize=8.5cm\epsfbox{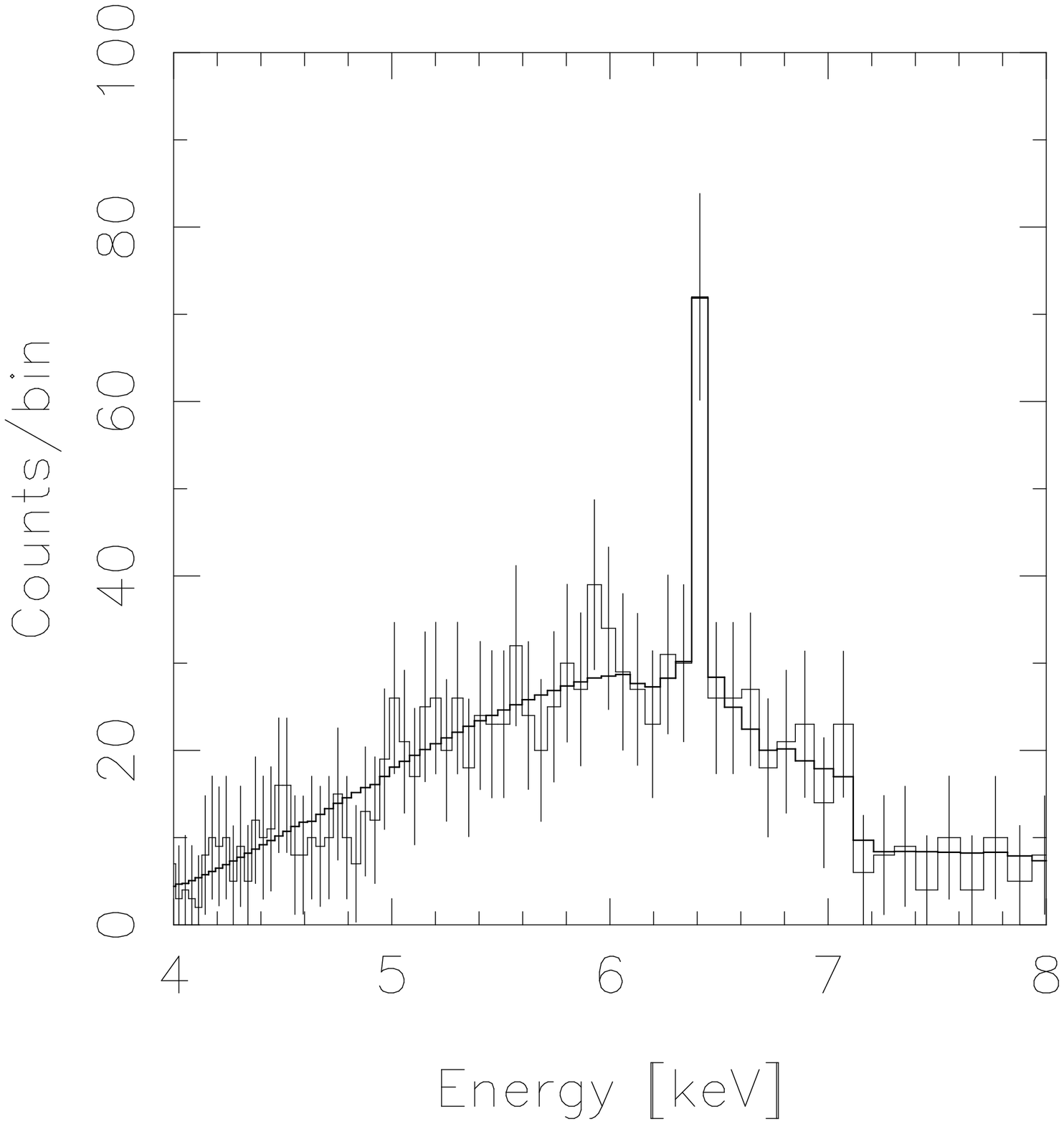}}
\figcaption{X-ray count spectrum of OAO 1657$-$415 from {\em
Chandra}/HETGS/HEG. The absorbed power-law 
spectral model shown has $N_{\rm H}=4.1\times10^{23}$~cm$^{-2}$,
photon index $\Gamma=1.0$, and normalization 0.046 photon cm$^{-2}$
s$^{-1}$ keV$^{-1}$ at 1~keV.  An Fe K emission line at 6.4~keV and an
Fe K absorption edge near 7.1 keV are clearly visible.}

\bigskip
\noindent
will yield the seventh dynamical measurement of the
neutron star mass in an X-ray binary.

Our X-ray spectral measurements indicate that OAO~1657$-$415 is
extremely absorbed.  However, though the optical extinction is also
large, it is an order of magnitude smaller than would be inferred
from the usual Galactic ratio of $N_{\rm H}/A_V = 1.8\times
10^{21}$~cm$^{-2}$~mag$^{-1}$ (Predehl \& Schmitt 1995).   
This probably indicates that most of the X-ray absorption is by gas
local to the binary, presumably fed by the companion's stellar wind,
and that this gas has a very low dust content (as expected for a
B~star wind).   The same phenomenon has been observed in the
well-known wind-fed X-ray pulsar Vela~X-1 (Sako et al. 1999).  Indeed, 
OAO~1657$-$415 and Vela X-1 ($d=1.9$ kpc) are very similar in most
respects.   Both pulsars are in eclipsing high-mass X-ray binaries
containing a B supergiant companion, have similar orbital parameters,
are accreting from their companion's stellar wind, and have similar X-ray
luminosities (for our derived distance, the 2--10~keV luminosity of
OAO 1657$-$415 is $3\times 10^{36}$ erg~s$^{-1}$).  Thus, we expect
that OAO~1657$-$415 should be a good candidate for studying the
ionized stellar wind of the B~supergiant with X-ray line spectroscopy,
as has been done with Vela X-1 (Sako et al. 1999; Schulz et
al. 2002). 

\acknowledgements{We thank Nicole van der Bliek, Robert Blum, Angel
Guerra, Sergio Pizarro, and the CTIO staff, as well as the ESO/NTT
service observing program, for their assistance and support.  We also
thank Kevin Tibbits for assistance in analyzing the {\em Chandra} data.
This work was supported in part by the Alfred P. Sloan Foundation.  It
was also supported by {\em Chandra} award GO1-2047X issued by the {\em
Chandra X-Ray Observatory Center}, which is administered by the
Smithsonian Astrophysical Observatory under NASA contract NAS8-39073.}

\end{document}